\documentclass[showpacs,prl,twocolumn,floatfix,reprint]{revtex4-2}
\usepackage{amsmath,graphicx,amssymb,hyperref}

\begin{document}

\title{Impact of the Ukrainian crisis on the global food security}

\author{Jean Cyrus de Gourcuff}
\affiliation{Universit\'e Paris-Saclay, CNRS, CEA, Institut de Physique Th\'{e}orique, 91191, 
	Gif-sur-Yvette, France}

\author{David Makowski}
\affiliation{Universit\'e Paris-Saclay, AgroParisTech, INRAE, Unit of Applied
  mathematics and computer science, France}

\author{Philippe Ciais}
\affiliation{Laboratoire des Sciences du Climat et de l’Environnement, LSCE/IPSL,
CEA-14 CNRS-UVSQ, Universit\'e Paris-Saclay, 91191 Gif-sur-Yvette, France
}
  
\author{Marc Barthelemy}
\email{marc.barthelemy@ipht.fr}
\affiliation{Universit\'e Paris-Saclay, CNRS, CEA, Institut de Physique Th\'{e}orique, 91191, 
	Gif-sur-Yvette, France}
\affiliation{Centre d'Analyse et de Math\'ematique Sociales (CNRS/EHESS) 54 Avenue de Raspail, 75006 Paris, France}

\begin{abstract}

   Using global wheat trade data and a network model for shock propagation, we study the impact of the Ukrainian crisis on food security. Depending on the level of reduction in Ukrainian wheat exports, the number of additional individuals falling under the minimum dietary energy requirement varies from 1 to 9 millions, and reaches about 4.8 millions for a $50\%$ reduction in exports. In the most affected countries, supply reductions are mainly related to indirect trade restrictions.
  
\end{abstract}


\maketitle

Wheat is at the core of food supply in many countries. This cereal is among the most widely cultivated crop in the world and covers about a fifth of food calories and protein requirements of the population \cite{erenstein2022global}. In 2020, more than half of the wheat exports were supplied by five countries, namely Russia, USA, Canada, France and Ukraine \cite{fao_data}. In 2019, Russia and Ukraine accounted for respectively $10.3\%$ and $3.4\%$ of the world wheat production and for $19.7\%$ and $10.0\%$ of the global wheat trades. Ukraine was the fifth wheat exporter in the world in 2020. Since wheat exports are concentrated in a handful number of countries, any disruption of the wheat trade could have major immediate humanitarian consequences worldwide. A better understanding of the propagation of production and export shocks could help policy makers and international organizations to design strategies to mitigate the effects of these shocks on food insecurity \cite{Alexander:2006,Dai:2011,Sheffield:2014,suweis2015resilience,Portner:2022}. Extreme weather has been one of the prevailing causes of production shocks in the past and their frequency and intensity are both expected to increase in the future. Besides climate-related shocks, trade can be disrupted by other factors, such as export restrictions and wars. During the last 30 years, wheat trade had not been strongly impacted by conflicts or political instabilities.

\begin{figure*}[ht!]
	\centering
	\includegraphics[width=0.8\textwidth]{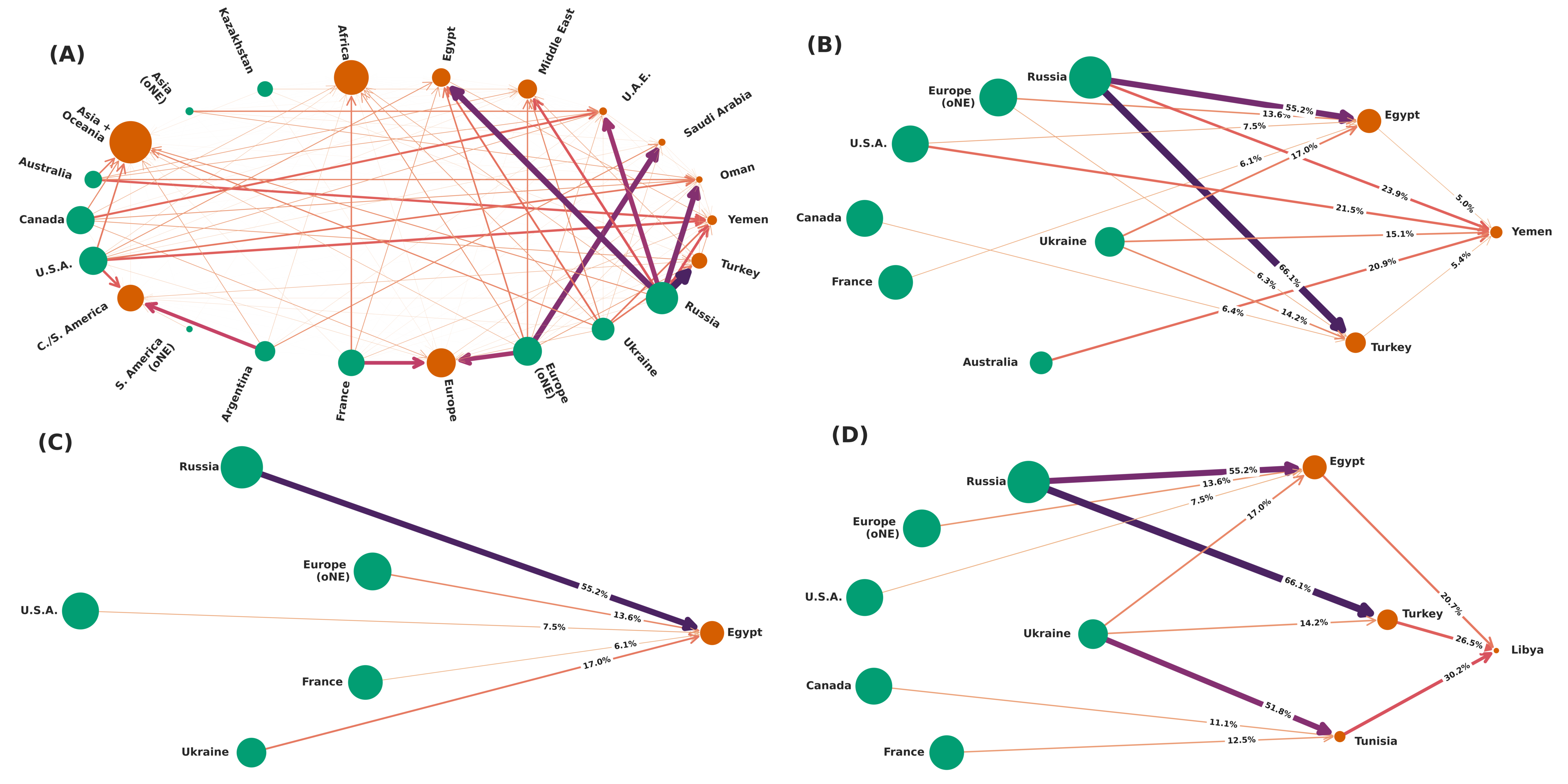}
	\caption{Wheat supply-chains of Yemen (A and B), Egypt (C) and Libya (D) from the 2019 GWTN. The extract process of these sub-networks consists of several  depth-first searches followed by filtering methods. (A) shows the complete network that results from this extraction process while (B), (C) and (D) have an extra step of node and edge filtering. The sub-network in (A) explains $100\%$ of Yemen total wheat imports and (B) only $91.95\%$, (C) explains $99.44\%$ of Egypt total wheat imports and (D) explains $77.37\%$ of Libya total wheat imports. The green (resp. orange) nodes are country/region that have a positive (resp. negative) wheat import/export balance and the node's size is proportional to that surplus (resp. deficit). On graphs (B), (C) and (D), the edge labels display the importing country's wheat import percentage this flow represents, while their color/width matches the absolute quantity traded. Aside from the net positive exporters, we distinguish 3 categories of countries: those that import almost exclusively directly from net positive exporters such as Egypt (C), those that import almost exclusively from intermediaries such as Libya (D) and those that import from both net positive and intermediaries such as Yemen (B).}
	\label{fig:examples}
\end{figure*}

But, since February 2022, the invasion of Ukraine by Russia has induced
substantial stock destructions and export restrictions with major consequences
on wheat trades \cite{Hassen}. Here, we analyze quantitatively the stability and resilience of a Global Wheat Trade Network (GWTN) - an important example of economic networks \cite{Fagiolo} - including $170$ nodes representing countries and between $4950$ (in 2010) and $5690$ (in 2019) directed and weighted edges representing yearly country-to-country amounts of traded wheat. The GWTN was calibrated using the wheat data provided by the Food and Agricultural Organization of the United Nations (FAO) from 2010 to 2019 \cite{fao_data, fao:foodBalance, fao:tradeMatrix}. We show in Fig.~\ref{fig:examples} several sub-networks of the 2019 GWTN that represent supply chains made of upstream flows for some specific countries. Among all the 173 countries in the graph and putting aside the 28 ($16\%$) net positive exporters, we distinguish 3 different types of countries according to their wheat importing behavior in the network. The first group (68 countries or $39\%$) comprises countries that import directly their wheat from an exporting source. The second group contains countries that
 only import from intermediaries (14 countries or $8\%$) and the third group is made of countries that display a mixed behavior (63 countries or $36\%$).

To assess the impact of a reduction in Ukrainian exports, we first assume a $50\%$ decline in Ukrainian wheat exports from the 2019 GWTN baseline, which represents a $3.8\%$ decline in the total quantity of wheat and wheat products reaching the world market. This order of magnitude of such a drop was previously estimated \cite{Lin}

\begin{figure*}[ht!]
  \includegraphics[width=0.8\textwidth]{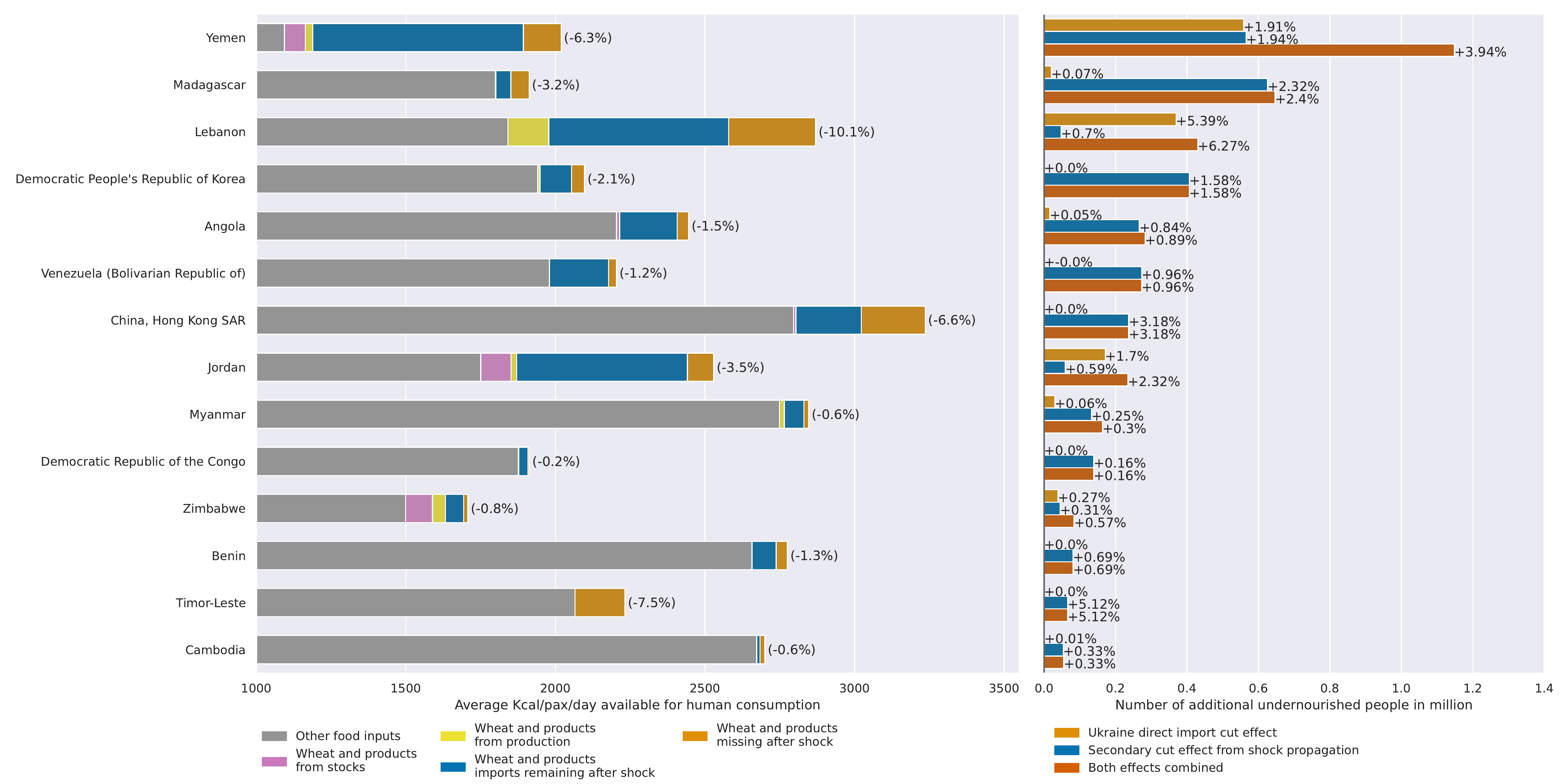}
    \caption{Average food supply and impact estimate in the case of a $50\%$
      reduction of Ukrainian wheat exports.
      \textbf{Left panel}: Average food supply (in Kcal/cap/day) available for the countries’ population. The
      share of food supply missing due to the shock is shown in orange and blue represents what remains after the shock. Only percentages above $3\%$ are shown. \textbf{Right panel}:
     Additional number of people who would fall below the
    undernourishment threshold if exports from Ukraine were to be
   reduced by $50\%$ with no mitigation mechanism. Numbers (in million) are given
   for direct (yellow), indirect (blue) and cumulative (orange)
   effects of the shock propagation. The percentages on the right
   side of the bars show theses numbers in terms of the percentage of the country's total population.}
    \label{fig:impact_shoc_ukrain_sorted_by_pop}
  \end{figure*}

This shock is then recursively propagated to other countries through a forward depth-first mechanism. We assume that a country absorbs an import loss if and only if it cannot be compensated by a reduction of its export. The import loss of a given country is distributed as much as possible to the country’s export partners according to a proportional strategy \cite{tamea2016global} (see Methods). As wheat is an essential component of the human diet, we assume that the wheat demand is inelastic \cite{Andreyeva:2010, Kumar:2011, Bentley:2022}. We do not consider any mitigation mechanism, as the objective here is to understand the underlying risks and cascading effects if nothing is done to mitigate the shock impact. Once the shock has propagated throughout the network, we compute the total amount $C_{tot}$ of food supply available in each country which in turn allows us to estimate the number of additional undernourished individuals in each country (for details, see Methods). Results show that the impact of this shock on the GWTN structure is substantial: the number of strictly importing countries increases from $39$ to $100$, the number of edges is divided by $1.7$ and the average quantity exported through a single trade relationship is increased by $60\%$. Because of the shock propagation, GWTN takes a shallow structure combining simple subnetworks (see Methods for examples of such subnetworks). The most impacted countries in terms of the number of individuals falling below the undernourishment threshold are presented in Fig.~\ref{fig:impact_shoc_ukrain_sorted_by_pop}, distinguishing the shares of reduced imports resulting from direct export cuts and from indirect propagation mechanisms. We note that over $90\%$ of Ukrainian wheat products are exported to Africa, Asia, or Middle-East, which is why all the most affected countries are from these regions. The most affected countries (i.e., Lebanon, Tunisia, Morocco, Jordan) are highly dependent on wheat for their food supply and moderately wealthy at best. Despite the fact that many of these countries do produce a lot of the wheat they consume, they remain highly vulnerable.  Maybe more surprisingly, countries with a small share of wheat in their food basket but with an overall low food supply are also vulnerable to the Ukrainian crisis. Typically, Madagascar, the Democratic People’s Republic of Korea or even Thailand are considered vulnerable even though relatively little wheat enters their food basket. In these countries, any small decrease in wheat supply puts hundreds of thousands of people at risk due to their low overall food supply.  Globally, our results show that a $50\%$ reduction in Ukrainian wheat export leads to a total of about an additional 6 million individuals that fall under the minimum dietary energy requirement. The countries concerned are affected at different levels, up to $6.3\%$ (for Lebanon) of their population. In the most vulnerable countries, the indirect trade restrictions are responsible for most wheat shortages. Considering export restriction ranging from $10\%$ to $100\%$, the number of affected individuals falling below minimum dietary requirements varies from 1 to 9 millions (see Methods).
  
Our data-driven network model fills a gap in understanding global food
security. It can be used to assess the impact of an export disruption, to identify the most vulnerable countries and assess different types of mitigation strategies. For example, a simple and appealing strategy could be to establish new trade partnerships with existing major exporting countries. This strategy has already been applied by Egypt; following the Ukrainian wheat export crisis, this country raised its wheat imports from Russian by $84\%$ in March-May 2022. However, this strategy could actually increase the vulnerability of the entire trading system. Indeed, with $80\%$ of world wheat exports currently coming from only 13 countries, increasing trade volumes with current partners may lead to greater dependence on a limited number of exporting countries. A better strategy would be to diversify the wheat production and storage areas in the world in order to increase the number of exporting countries and decrease the dependence of the most vulnerable importing countries on a few players.


\section{Material and methods}

\subsection{GWTN data}

All the data comes from the Food and Agricultural Organisation of the United Nations open source databases \cite{fao_data,fao:foodBalance, fao:tradeMatrix, fao:foodSecIndic}. In particular, we used the following databases:
\begin{itemize}
\item{} {\it Food Balances (2010-)} contains aggregated data on countries yearly domestic utilisation details for every aggregated group of agricultural commodity (eg: wheat and products) and for every year from 2010 to 2019. The calories supply per capita obtained from these commodity and the population sizes are also provided. The data was revised, completed and made coherent following the new DAO's FBS methodology as described on the data download page.
  \item{} {\it Detailed Trade Matrix} contains all agricultural commodity trades between countries from 1986 to 2020 with a yearly resolution. The data is not reconciled, meaning that countries may have reported contradictory information. We therefore had to reconcile this dataset, following the main idea that in case of a conflict, the last word is given to the importer, as suggested by the FAO guidelines on the download page. More information is available in the Methods.
    \item{} {\it Suite of Food Security Indicators} contains series of food security indicators for every country and every years from 2000 to 2001. Only the Minimum Dietary Energy Requirement and the Coefficient of Habitual Caloric Consumption were retained.
    \end{itemize}

    The data sets were all downloaded in their entirety on the $7^{th}$ of July, 2022 from the dedicated FAO pages \cite{fao:foodBalance, fao:tradeMatrix, fao:foodSecIndic}.

\subsection{GWTN Structure}
    
We analyzed the stability and resilience of the Global Wheat Trade Network (GWTN) following
the same lines as in \cite{serrano2003topology, garlaschelli2005structure, serrano2007patterns, gutierrez2021analysis}.

The GWTN is modeled by a set of vertices $1\leq j \leq N$ embedding the countries and a set of edges $a_{ij}$ and weights $w_{ij}$, $1 \leq i,j \leq N$ where $a_{ij} = 1$ if and only if country $j$ imports $w_{ij} > 0$ tons of wheat from country $i$.
We define the in-degree of country $j$ as $k^{in}_{j} = \sum\limits_i a_{ij}$ and its weighted counterpart $I_j = \sum\limits_i w_{ij}$ - the sum of all wheat and wheat products imports of country $j$. Similarly, define the out-degree of country $j$ as $k^{out}_{j} = \sum\limits_i a_{ji}$ and its weighted counterpart $E_j = \sum\limits_i w_{ji}$ - the sum of all wheat and wheat products exports of country $j$.

The GWTN, as exported from the FAO database \cite{fao:foodBalance, fao:tradeMatrix}, is a $\sim$170 countries network, and the number of reported trade partnerships (edges) went from roughly 4950 in 2010, to 5690 in 2019, making its density going from 0.17 to .19 during the same period. The network's components sizes are quite stable over the years, with a giant strongly connected component of around 130 countries (with average distance in this component equal to $1.95$) and an out-component (pure importers) of around 40 countries.

The weights $(w_{ij})_{1 \leq i, j \leq N}$ distribution follows a power law of parameter $\alpha \approx 1.27$ and features stability over the years (fig. SI-1). While the in-degrees $(k^{in}_j)_{1 \leq j \leq N}$ display a peaked distribution (fig. SI-2), the out-degrees $(k^{out}_j)_{1 \leq j \leq N}$ are distributed along a flat-tailed distribution (fig. SI-3). Finally we note that the diameter of the network, defined as the maximum shortest directed path length is of $log(N) \approx 5$. The average in-degree of this network is about $33$ showing a high level of interconnectedness.

\begin{figure}[h]
	\centerline{
		\includegraphics[scale=.30]{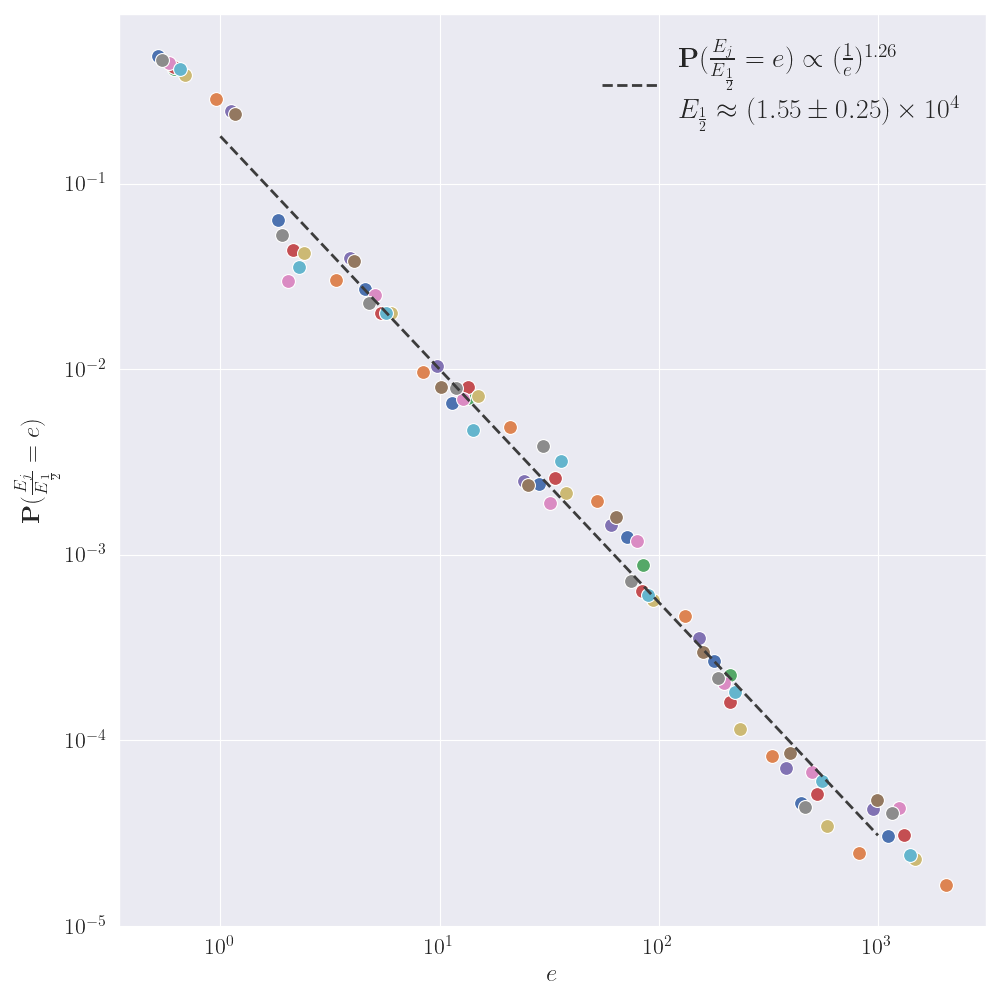}}
	\caption{Probability distribution of the total countries normalized exports $\frac{E_{j}}{E_{\frac{1}{2}}} = \frac{1}{E_{\frac{1}{2}}}\sum\limits_i w_{ji}$ for each year from 2010 to 2019 (differentiated by the colors). The collapse between each year $t$ is obtained by normalizing by $E_{\frac{1}{2}}(t)$, the median of the $E_j(t)$ for $2010 \leq t \leq 2019$. Once normalized, the data were binned using a power scale of parameter $\alpha=2.5$.}
	\label{fig:out_strength}
\end{figure}

Figure \ref{fig:out_strength} shows the distribution of the countries total exports $(E_{j}(t))_{1 \leq j \leq N}$ for all ten years $2010 \leq t \leq 2019$. Since the total volume of exports increased from 2010 to 2019, we normalized each data sample $(E_{j}(t))_{1 \leq j \leq N}$ by the median of the samples $E_{\frac{1}{2}}(t)=Med(\{E_{j}(t), \quad 1 \leq j \leq N\})$ which resulted in the distributions collapsing together. We clearly have a fat tailed distribution, which entails that most of the exported wheat comes from a handful number of countries, and to such proportions that the quantities exported by the little exporters are negligible in first approximation.

From the import point of view, we zoom on the countries that have a negative import/export balance ($E_j < I_j$) and import from at least two different countries. Among these 130 countries, half of them import more than half of their wheat from only one country, and around 85\% of them get at least 30\% of their imported wheat from one trade partner. Another way to look at it is to look at the Gini coefficients $(G_j)_{1\leq j \leq N}$ for each country computed over their imports - that is
\begin{align}
  G_j = Gini(\{ w_{ij}|\; 1\leq\ i \leq N, \; w_{ij} > 0\})
\end{align}
These coefficients range from .7 to 1 and due to size of the samples, are most likely downward biased \cite{Deltas:2003}. All in all, if the distributions showcased in fig. SI-2 may let us think that countries have managed to create diversified sources of wheat income, the above facts show quite the opposite. Negatively balanced countries rely for the most part on a very few countries to secure their food supply.

We then defined the distance between two connected nodes $i$ and $j$ as $\delta^{w}(i, j) = 1 - \frac{w_{i, j}}{I_j}$ if $a_{i, j} = 1$. In the case of two connected countries, this distance is minimal when country $j$ imports all of its wheat from country $i$, and conversely, converges towards 0 the smaller the imports from country $i$ are relatively to country $j$ imports. This metric can be understood as a measure of the efficiency of a link, or from the importer point of view, as a measure of the import dependency of a country with respect to the other. From there, we define the shortest directed weighted path length between $i$ and $j$ as $d^{w}_{ij} = \frac{1}{d_{ij}}\sum\limits_{(u, v)\in g(i,j)}\delta^{w}_{ij}(u, v)$ where $g(i, j)$ is the shortest weighted directed path relatively to the metric $\delta^{w}$. $d^{w}$ ranges from $0$ to $1$ and measures how close country $i$ is from country $j$ in the network relatively to the metric $\delta^{w}$. Fig. SI-5 displays a violin plot of the computed $d^{w}_{ij}(t)$, $2010 \leq t \leq 2019$ for all three sets $d_{ij} = 1$, $d_{ij} = 2$ and $d_{ij} \in \{3, 4, 5\}$. The area of the violins is proportional to the number of sample they embed and we aggregated the last three sets so that the samples size are within the same order of magnitude. The results show us that if the shortest directed path lengths are small enough so that we may talk of a small world, the weighted point view overturns this conclusion. From the tails of the distribution, long and decreasing in size as $d_{ij}$ increases, we can assert that only a few of the trade routes are truly significant and that those that are significant are more likely to be those with the fewest intermediaries.  

\begin{figure}[h]
	\centerline{
		\includegraphics[scale=.3]{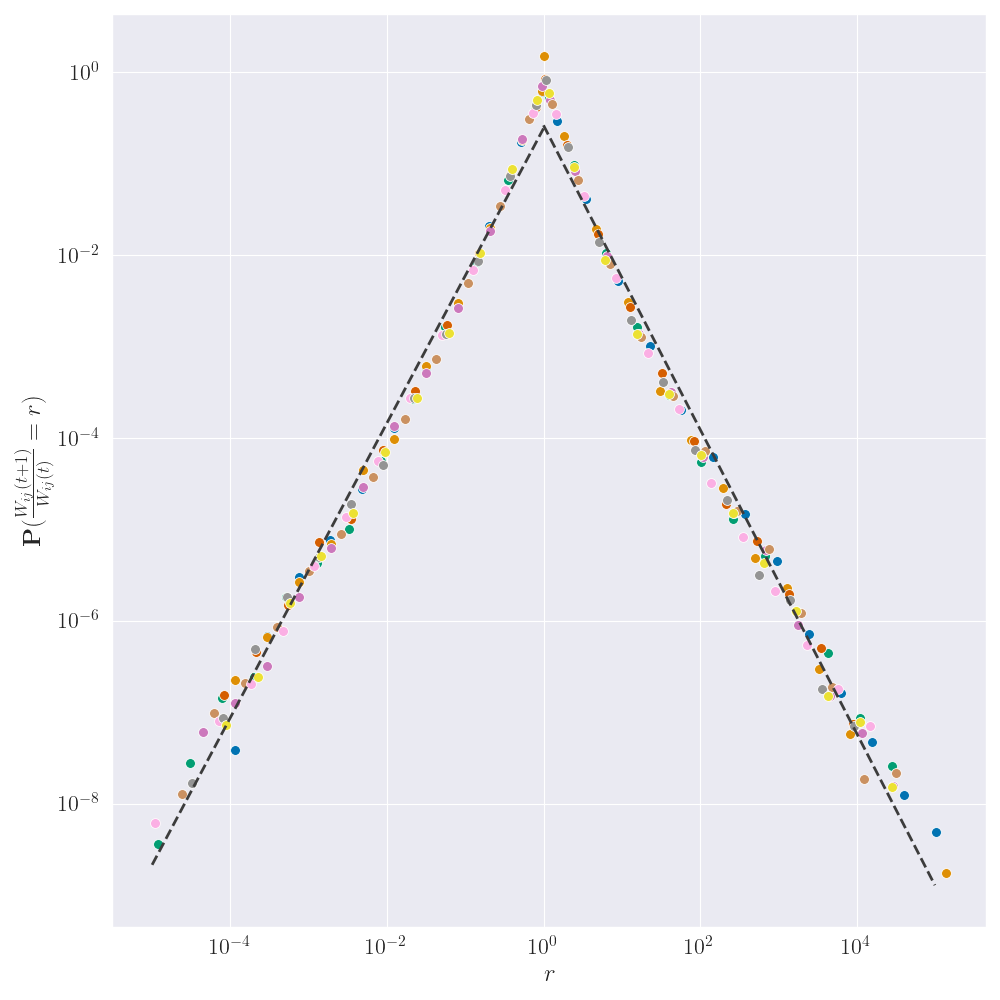}}
	\caption{Probability distribution of the consecutive weight ratio $r_{ij}(t) = \frac{w_{ij}(t+1)}{w_{ij}(t)}$  for each year from 2010 to 2019 - each year corresponding to a different color. The data were binned using a power scale of parameter $\alpha=2.55$. The dashed lines have a slope of roughly 1.6 so that $\mathbf{P}(\frac{W_{ij}(t+1)}{W_{ij}(t)} = r) \propto e^{-1.6|ln(r)|}$.}
	\label{fig:log_weight_ratio}
\end{figure}

As a mean to understand the flows and trade partnerships dynamics, we define the ratio $r_{ij}(t)$ as the ratio between the traded quantities during a year $t$ and the year $t+1$, if country $j$ imported wheat from country $i$ during both years, that is $r_{ij}(t) = \frac{w_{ij}(t + 1)}{w_{ij}(t)}$. Figure \ref{fig:log_weight_ratio} shows the distribution of these ratios for each year from 2010 to 2019. The distribution of the ratio is strikingly stable over the years. Namely, for all $t$ we have that the distribution of a link's weight ratio $r_{ij}(t) = \frac{w_{ij}(t+1)}{w_{ij}(t)}$ - given that it exists at both years $t+1$ and $t$, is proportional to $r \xrightarrow{} e^{-(1+\beta)|ln(r)|}$ with $\beta \simeq 0.6$.

Thus, the GWTN displays at first glance the typical features of a scale-free small-world network \cite{Watts:1998, Barabasi:2003}, as many real world systems are, it is actually structurally and dynamically speaking much closer to a very shallow forest, where the wheat flow cascades from a few roots, namely the biggest exporters. The surprising feature of this network is that the vast majority of the exports and production come from a handful of countries. Some countries naturally act as trade platforms as they are a step of important trade routes. Others may import raw materials to process them and export them again. Finally, to ensure food security for a population, a reasonable strategy is to multiply the trade partnerships, even if most of the imports must be re-exported to keep a positive commercial balance, so that the risk of a food supply scarcity from a trade failure is reduced.
\begin{figure}[ht!]
    \centerline{
    \includegraphics[scale=.30]{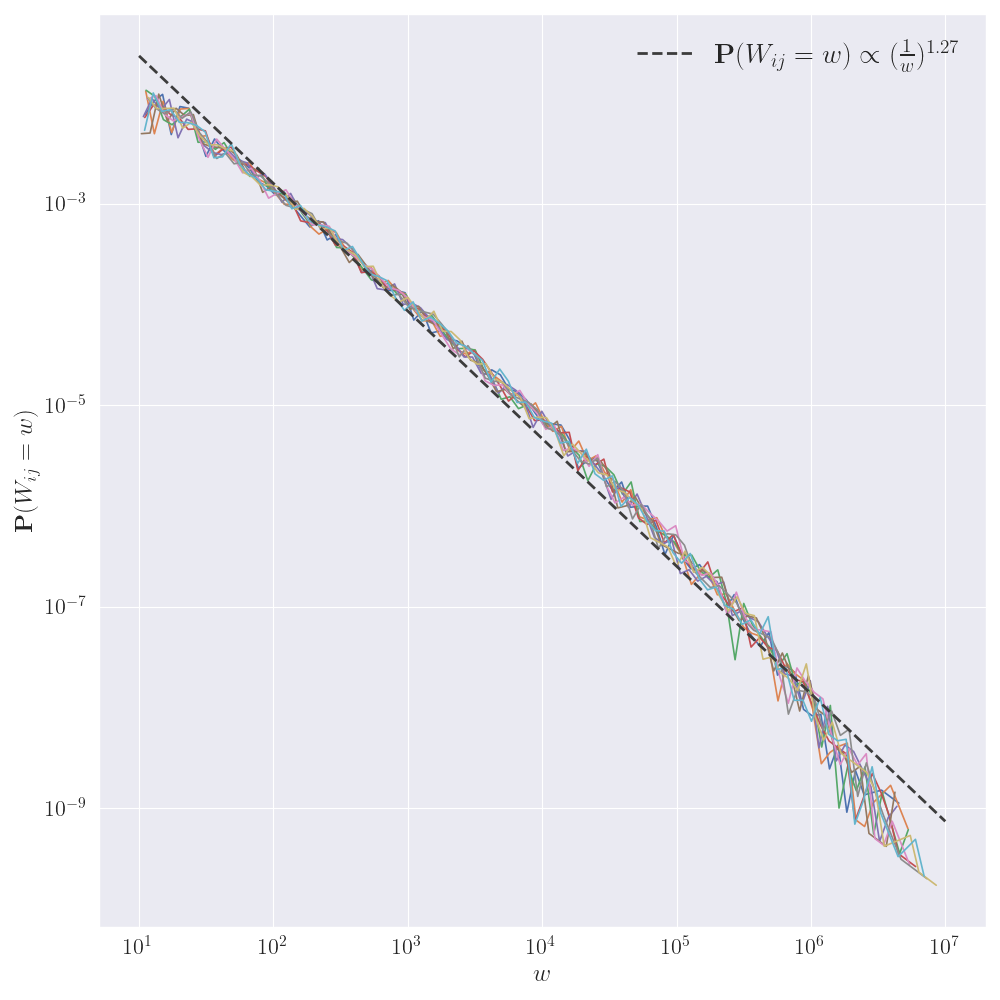}}
    \caption{Probability distribution of the weights $w_{ij}(t)$ for each year $2010 \leq t \leq 2019$ (differentiated by the colors). The data were binned using a power scale of parameter $\alpha=1.16$.}
    \label{fig:weights_distrib}
\end{figure}
\begin{figure*}[ht!]
    \centerline{
    \includegraphics[scale=.4]{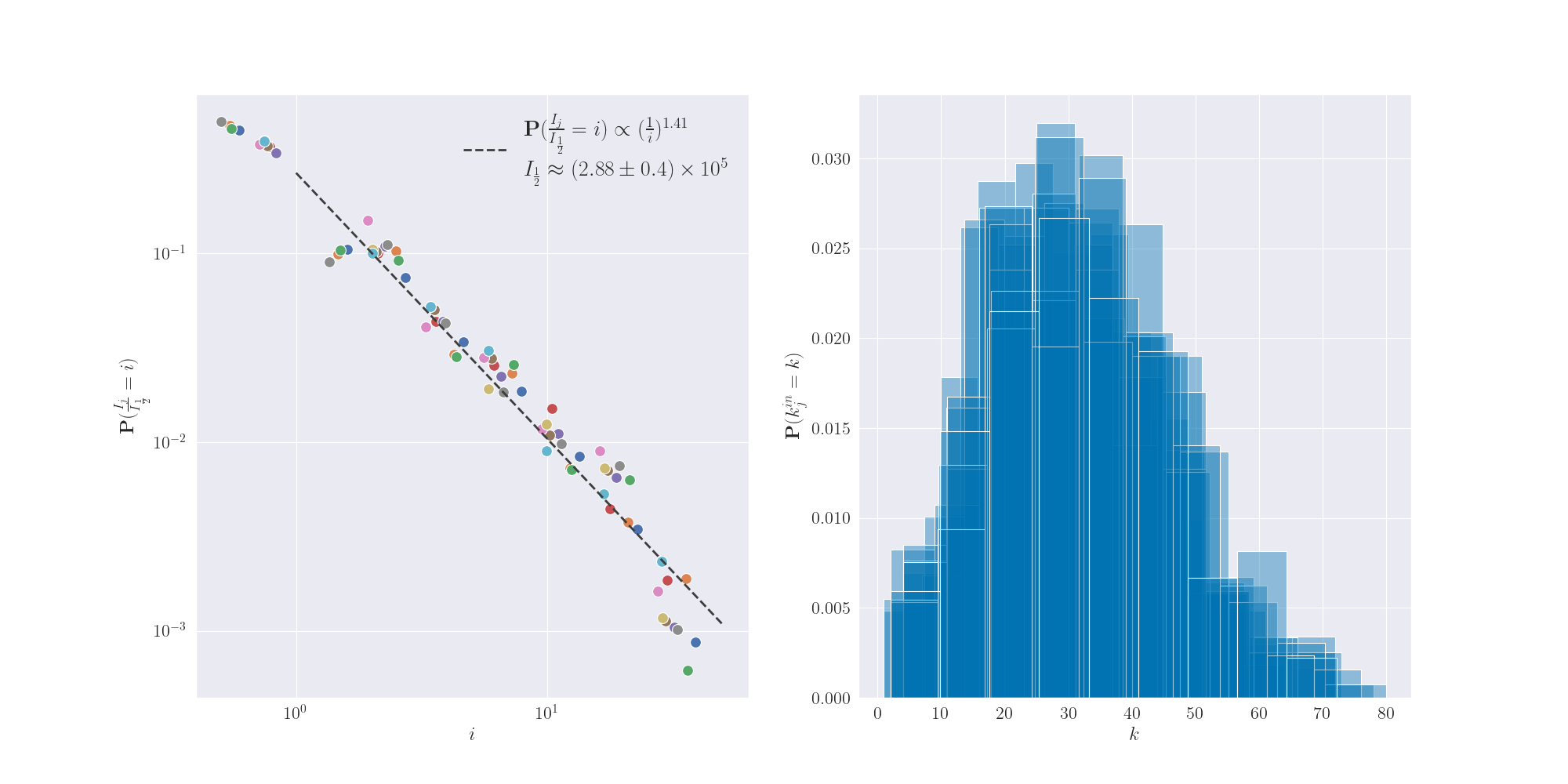}}
    \caption{\textbf{Left panel}: Probability distribution of the total countries normalized inports $\frac{I_{j}}{I_{\frac{1}{2}}} = \frac{1}{I_{\frac{1}{2}}}\sum\limits_i w_{ij}$ for each year from 2010 to 2019 (differentiated by the colors). The collapse between each year $t$ is obtained by normalizing by $I_{\frac{1}{2}}(t)$, the median of the $I_j(t)$ for $2010 \leq t \leq 2019$. Once normalized, the data were binned using a power scale of parameter $\alpha=1.7$. \textbf{Right panel}: Probability distribution of the countries in degrees $k^{in}_j$ from 2010 to 2019.}
    \label{fig:in_strength}
\end{figure*}
\begin{figure}[ht!]
    \centerline{
    \includegraphics[scale=.30]{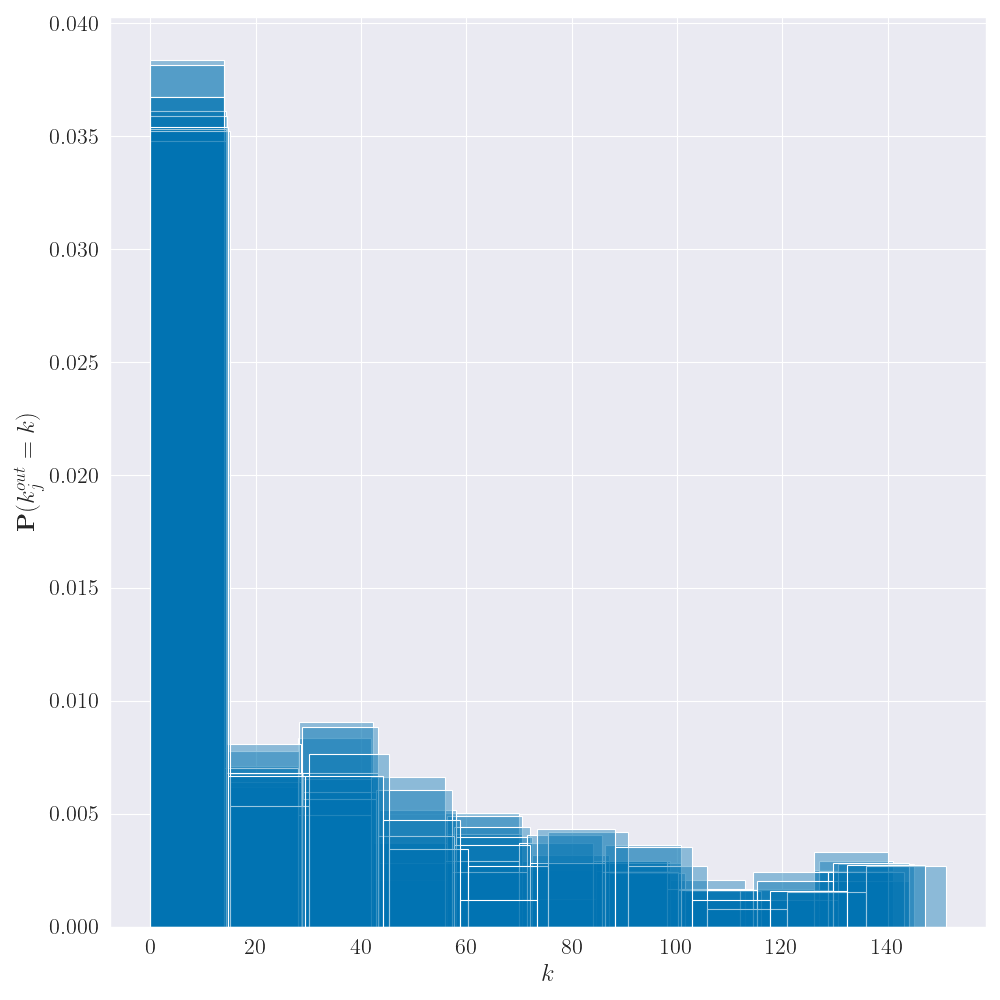}}
    \caption{Probability distribution of the countries out degrees $k^{out}_j$ from 2010 to 2019.}
    \label{fig:out_degree}
\end{figure}
\begin{figure}[ht!]
    \centerline{
    \includegraphics[scale=.3]{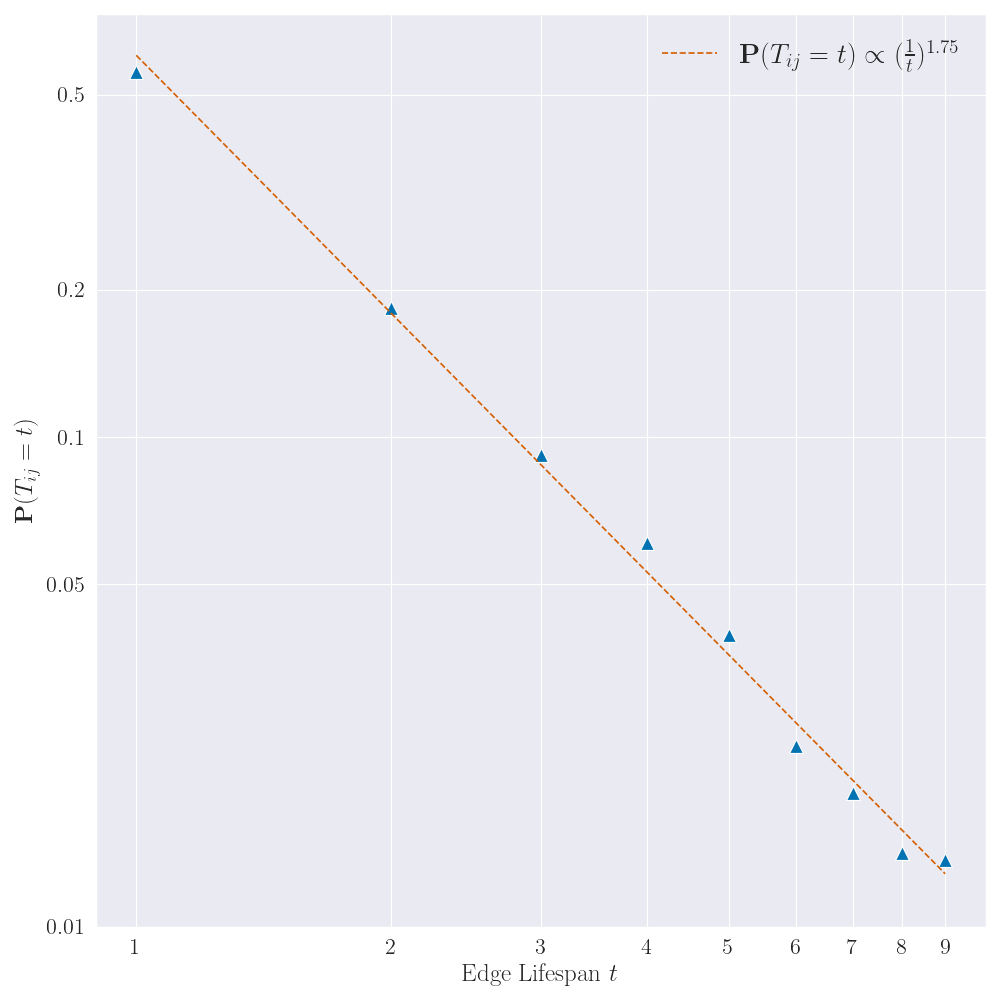}}
    \caption{Probability distribution of trade partnerships lifespans from 2010 to 2019.}
    \label{fig:link_lifespan}
\end{figure}
\begin{figure}[ht!]
    \centerline{
    \includegraphics[scale=.35]{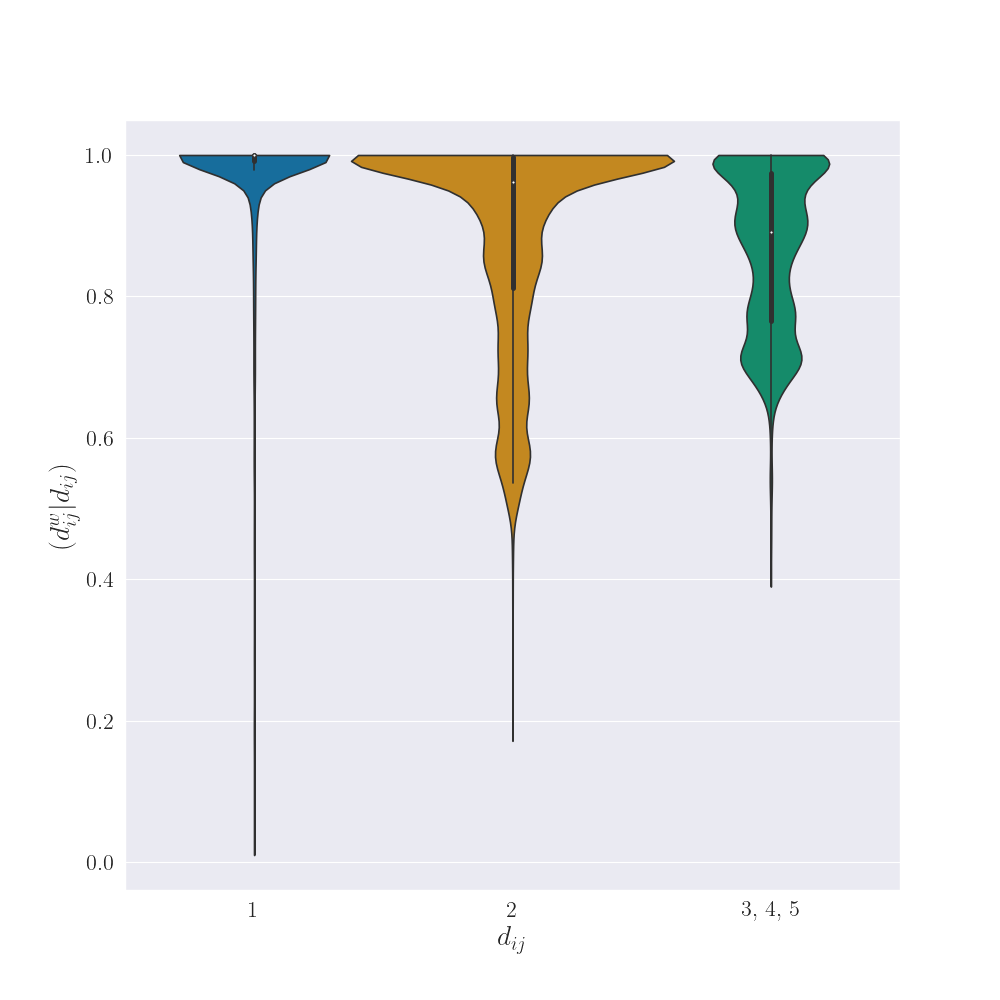}}
    \caption{Distributions of the Shortest Weighted Directed Path Lengths $d^{w}_{ij}$ given the Shortest Directed Path Length ${d_{ij}}$ in year 2019. The Shortest Weighted Directed Path between two nodes $i$ and $j$ is defined as $d^{w}_{ij} = \frac{1}{d_{ij}}\sum\limits_{a_{uv}\in g_{i \rightarrow j}}1 - \frac{w_{uv}}{I_{v}}$, with $g_{i \rightarrow j}$ the shortest weighted path from $i$ to $j$ and $\forall v, I_{v} = \sum \limits_k w_{kv}$ the sum of all country $v$ imports. The violins shapes show the distribution (the areas are scaled relatively to the number of occurrence in each sub-category) while is displayed inside of each violin the corresponding box-plot.}
    \label{fig:shortests_paths}
  \end{figure}

\subsection{Shock propagation in the network}

We choose a source (in this study Ukraine) and reduce its export by a given percentage. We propagate this shock through the network using a redistribution strategy (see below). Once this is done, we obtain the total supply of food $C'_{tot}$ available in each country (see below). From this quantity we can then compute the number of individuals below the minimum dietary requirement.

\subsubsection{Redistribution strategy}

\begin{figure}[ht!]
  \centering
	\includegraphics[scale=.35]{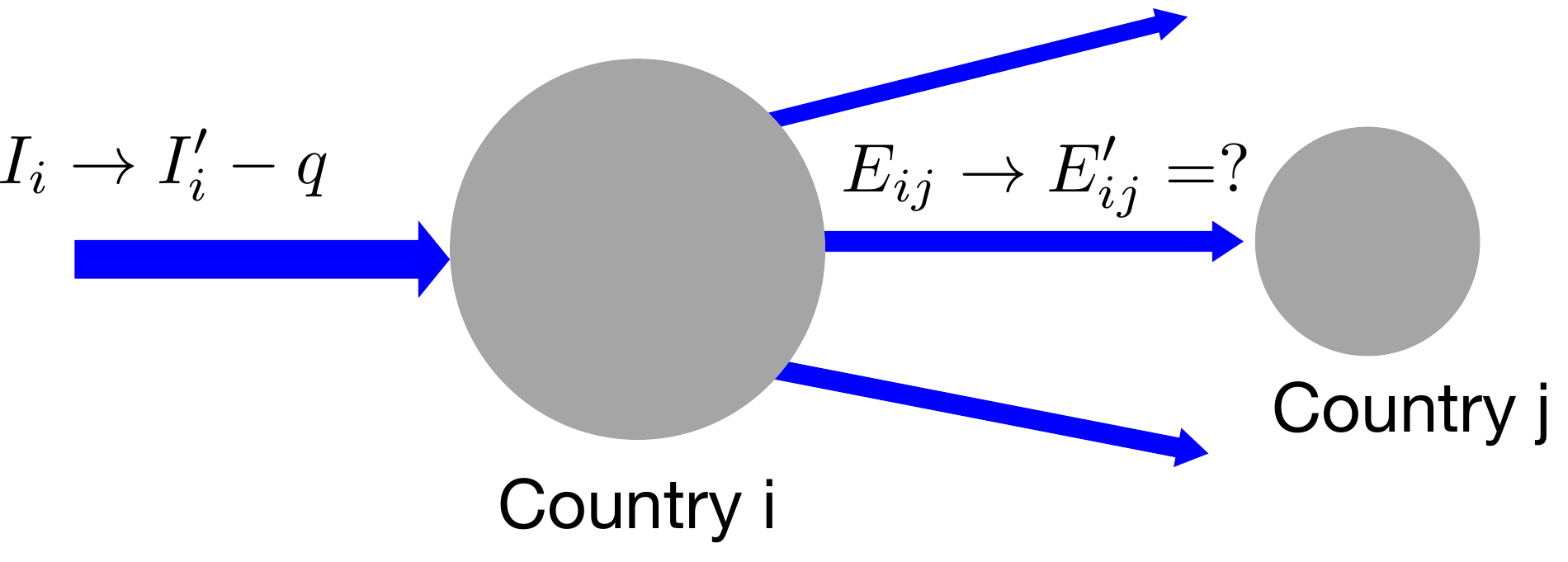}
        \caption{Illustration of the shock propagation. The import of a country is reduced by a quantity $q$ which
          will impact its export to other countries.}
\label{fig:reduc}
\end{figure}

We assume that the import of a country $i$ is reduced by a quantity $q$ (see Fig. \ref{fig:reduc}). 
\begin{align}
	I_i\to I'_i=I_i-q
\end{align}

Here, we will choose the proportional strategy as suggested in previous works \cite{tamea2016global}.  For this strategy all export will suffer from a decrease proportional to their share of the total export. In other words, the export from country $i$ to country $j$ will then be
\begin{align}
	E_{ij}\to E'_{ij}=E_{ij}-q\frac{E_{ij}}{\sum_l E_{il}}
\end{align}

We explored three strategies - proportional distribution of the import loss to all recipients, uniform distribution, and inequality reproductive distribution. We found by simulating shocks on the twenty biggest exporters, that the shock impacts were statistically independent from the redistribution strategy choice.

\subsection{Available Supply of Calories}

In what follows, we provide details on how we estimate the amount of food available per person per year, before and after an export/production shock for a given product occurs.

\subsubsection*{Terminology}
We clarify here the notations used. Quantities are classified according to the database from which they are extracted.

\paragraph{Indices and time variable}
As for the indices, we use
\begin{itemize}
    \item $1\leq i,j \leq N_{country}$ to designate countries
    \item $\alpha_k$, $1\leq k \leq N$ to designate commodities
    \item $2010 \leq t \leq 2019$ refers to the year
    \item $t + dt$ refers to the time right after a shock has occurred in year $t$. (The shock propagation is supposed to be instantaneous).
\end{itemize}

\paragraph{Food balance}
The data extracted from the FAO food balance sheets \cite{fao:foodBalance} are:
\begin{itemize}
    \item $C_i^{\alpha}(t)$ - Number of calories available for human consumption from commodity $\alpha$ (Kcal/capita/day)
    \item $C^{tot}_i(t)$ - Total number of calories available for human consumption (Kcal/capita/day)
    \item $Q_i^{\alpha}(t)$ - Quantity of food made available for human consumption (tons per year)
    \item $\delta S_{i}^{\alpha}(t)$ - Changes in stocks (tons per year)
    \item $P_i^{\alpha}(t)$ - Production of commodity $\alpha$ (tons per year)
    \item ${Pop}_i(t)$ - Population (\# of inhabitants)
\end{itemize}
from which we compute:
\begin{itemize}
    \item $f_i^{\alpha}(t) = \frac{C_i^{\alpha}(t)}{C^{tot}_i(t)}$ - The share commodity $\alpha$ accounts for in the average food basket (dimensionless)
    \item $\delta^- S^{\alpha}_i(t) = - min(\delta S^{\alpha}(t), 0)$ - The quantity of commodity $\alpha$ from the stocks that can be added to the food balance.
\end{itemize}

\paragraph{Trade matrix}
The data extracted from the FAO detailed trade matrix \cite{fao:tradeMatrix} are:
\begin{itemize}
    \item $w_{ij}^{\alpha}(t)$, the quantity of commodity $\alpha$ country $i$ exported to country $j$ (tons)
\end{itemize}
from which we compute
\begin{itemize}
    \item $I_i^{\alpha}(t) = \sum\limits_j w_{ji}^{\alpha}(t)$ - the total volume of commodity $\alpha$ imported by country $i$ (tons per year)
    \item $E_i^{\alpha}(t) = \sum\limits_j w_{ij}^{\alpha}(t)$ - the total volume of commodity $\alpha$ exported by country $i$ (tons per year)
\end{itemize}

\paragraph{Inconsistencies within and between FAO's databases}
First, the trade data are not reconciled, therefore the trade matrix data are incoherent.
Second, many figures from the FAO's databases are estimations. Thus, there are some inconsistencies between the trade matrix data and the food balance data.

To circumvent this issue, we introduce error terms $\Delta E_i^{\alpha}(t)$ and $\Delta I_i^{\alpha}(t)$ with values before the shock $\hat{I}^{\alpha}_i(t) - I^{\alpha}_i(t)$ and $\hat{E}^{\alpha}_i(t) - E^{\alpha}_i(t)$, where $\hat{I}$ and $\hat{E}$ are the imports and exports values as reported in the food balance database.

\paragraph{Food security indicators}
The data extracted from the FAO food security indicators \cite{fao:foodSecIndic} are:
\begin{itemize}
    \item $M_i(t)$ - The minimum dietary energy requirement (MDER) (Kcal/capita/day)
    \item $cv_i(t)$ - The coefficient of variation of habitual caloric consumption (dimensionless)
\end{itemize}

from which we compute
\begin{itemize}
    \item $\widetilde{C_i^{tot}}(t) = \frac{C_i^{tot}(t)}{M_i(t)}$ - the normalized total number of calories available for human consumption (dimensionless)
\end{itemize}

\paragraph{Total quantity of commodity available}
As we are looking at worst case scenarios, the value of interest is the maximum quantity of commodity $\alpha$ that a country may mobilize. This quantity is given by:

\begin{equation}\label{D_de}
    D_i^{\alpha}(t) = I_i^{\alpha}(t) + P_i^{\alpha}(t) + \delta^- S_i^{\alpha}(t) + \Delta I_i^{\alpha}(t)
\end{equation}

\bigbreak
From now on the indices are omitted when there is no ambiguity.

\subsubsection*{Scope and assumptions}

After a shock, we have discounted values of the $w_{ij}$, $1\leq i,j \leq N_{country}$ for an export shock and discounted values of $(P_i^{\alpha})_{1\leq i \leq N_{country}}$ for a production shock. So for each country we know $I(t + dt)$ and $P(t + dt)$ and we want to estimate $C^{tot}(t + dt)$.

\paragraph{$\Delta I^{\alpha}$ variation}

As there is no mean to know how the import quantity from the error term is affect by the shock (the shock propagates at a local level while the error term is computed at a global level), and to keep things as simple as possible, we assume that the magnitude of the import error term diminishes proportionally to the ratio $\frac{I^{\alpha}(t+dt)}{I^{\alpha}(t)}$. That is

\begin{equation}\label{delta_i}
    \Delta I^{\alpha}_i(t + dt) = \frac{I^{\alpha}(t+dt)}{I^{\alpha}(t)}\Delta I^{\alpha}_i(t)
\end{equation}

\paragraph{$\delta^-S^{\alpha}$ variation}
We only have information about stocks variation. Therefore the only reasonable hypothesis is that
\begin{equation}\label{stock_var}
    \delta^-S^{\alpha}_i(t+dt) = \delta^-S^{\alpha}_i(t)
\end{equation}

\paragraph{Relation between $C^{\alpha}$ and $Q^{\alpha}$, conversion from tons to Kcal}
It is not possible to define a constant $\rho^{\alpha}$ such that
\begin{align*}
    \forall i, \; C_i^{\alpha} = \rho^{\alpha} Q^{\alpha}_i
\end{align*}
Indeed, in food balance sheets, $Q$ and $C$ include the raw material, but also all its derivatives.
And since the breakdown of commodity consumption varies from country to country, the conversion rate from tons to Kcal is not universal.
In that regard, we consider that $\frac{C_i^{\alpha}}{Q^{\alpha}_i}$ stands as the conversion rate for commodity $\alpha$, country $i$ and year $t$.

So that in the end,
\begin{equation}\label{tons_to_kal_conversion}
    C^{\alpha}(t + dt) = \frac{C^{\alpha}(t)}{Q^{\alpha}(t)}Q^{\alpha}(t + dt)
\end{equation}

\paragraph{Worst case scenario}
As our study's goal is to understand the worst case scenario of a shock, we assume that as a country is impacted by a shock on commodity $\alpha$, it redirects all of its available resources of that commodity to feed its population. Namely, we suppose that
\begin{equation}\label{q_equals_d}
    Q^{\alpha}_i(t+dt) = \left\{
        \begin{array}{lll}
            Q^{\alpha}_i(t) & \mbox{if}& D^{\alpha}_i(t+dt) \geq  Q^{\alpha}_i(t)\\
            D^{\alpha}_i(t+dt) & \mbox{otherwise.} &
        \end{array}
    \right. 
\end{equation}

\subsubsection*{Total number of calories available for the population}

We have for a shock on commodity $\alpha$:
\begin{align*}
    C^{tot}(t + dt) &= \sum\limits_{\beta}C^{\beta}(t + dt) \\
    &= \sum\limits_{\beta \neq \alpha}C^{\beta}(t) + C^{\alpha}(t + dt) \\
    &= C^{tot}(t) + C^{\alpha}(t + dt) - C^{\alpha}(t) \\
                    &= C^{tot}(t) + \frac{C^{\alpha}(t)}{Q^{\alpha}(t)}Q^{\alpha}(t + dt) - C^{\alpha}(t)
      \label{tons_to_kal_conversion}
\end{align*}
From there, if $D^{\alpha}_i(t+dt) \geq  Q^{\alpha}_i(t)$, then $C^{tot}(t+dt) = C^{tot}(t)$.
Otherwise, 
\begin{align*}
  C^{tot}(t + dt) &= C^{tot}(t) + \frac{C^{\alpha}(t)}{Q^{\alpha}(t)}(D^{\alpha}(t + dt) - Q^{\alpha}(t))\\
                  &= C^{tot}(t) + C^{\alpha}(t)[\frac{1}{Q^{\alpha}(t)}(I^{\alpha}(t+dt) \\
  &+ P^{\alpha}(t+dt) + \delta^- S^{\alpha}(t+dt) \\
    +& \Delta I^{\alpha}(t+dt))  - 1]\\
                  &= C^{tot}(t)(1 + f^{\alpha}(t)[\frac{1}{Q^{\alpha}(t)}(I^{\alpha}(t+dt) \\
  &+ P^{\alpha}(t+dt) + \delta^- S^{\alpha}(t)\\
    +& \frac{I ^{\alpha}(t+dt)}{I ^{\alpha}(t)}\Delta I^{\alpha}(t))- 1]) \\
\end{align*}

Here we note that the quantity $Q^{\alpha}(t)$ acts as the demand for commodity $\alpha$ so that the country can provide $C^{\alpha}$ to its population. Let us denote the constant for the current year $t$, $Q^{\alpha} = Q^{\alpha}(t)$ and
\begin{align*}
  x_i^{\alpha} &= \frac{D^{\alpha}_i(t+dt)}{Q_i^{\alpha}}\\
               &= \frac{1}{Q_i^{\alpha}}[I_i^{\alpha}(t+dt) + P_i^{\alpha}(t+dt) \\
  &+ \delta^- S^{\alpha}_i(t) + \frac{I ^{\alpha}_i(t+dt)}{I ^{\alpha}_i(t)}\Delta I_i^{\alpha}(t)] \\
\end{align*}
the fraction of the demand for commodity $\alpha$ that is met by the production, the imports and the stocks (dimensionless).

So that, 
\begin{align*}
    C^{tot}(t+dt) =  C^{tot}(t)(1 + f^{\alpha}(t)(x^{\alpha} - 1))
\end{align*}
and
\begin{equation*}
    \widetilde{C^{tot}}(t+dt) =  \widetilde{C^{tot}}(t)(1 + f^{\alpha}(t)(x^{\alpha} - 1))
\end{equation*}

To free us from heavy notations, we denote
\begin{align*}
    \widetilde{C^{tot}} &= \widetilde{C^{tot}}(t + dt) \quad \widetilde{C^{tot}_0} = \widetilde{C^{tot}}(t)  \quad f^{\alpha} = f^{\alpha}(t)\\
\end{align*}
so that:
\begin{equation*}
    \widetilde{C^{tot}} = \widetilde{C^{tot}_0}(1+f^{\alpha}(x^{\alpha} - 1))
\end{equation*}

At the end of the day:
\begin{equation}\label{F_a}
    \widetilde{C^{tot}} = \left\{
    \begin{array}{lll}
        \widetilde{C^{tot}_0}(1+f^{\alpha}(x^{\alpha} - 1)) & \mbox{if} &  x^{\alpha}  < 1 \\
        \widetilde{C^{tot}_0} & \mbox{otherwise.} 
    \end{array}
    \right.
\end{equation}

\subsection{Variable export drop}

\begin{figure}[ht!]
	\includegraphics[width=0.4\textwidth]{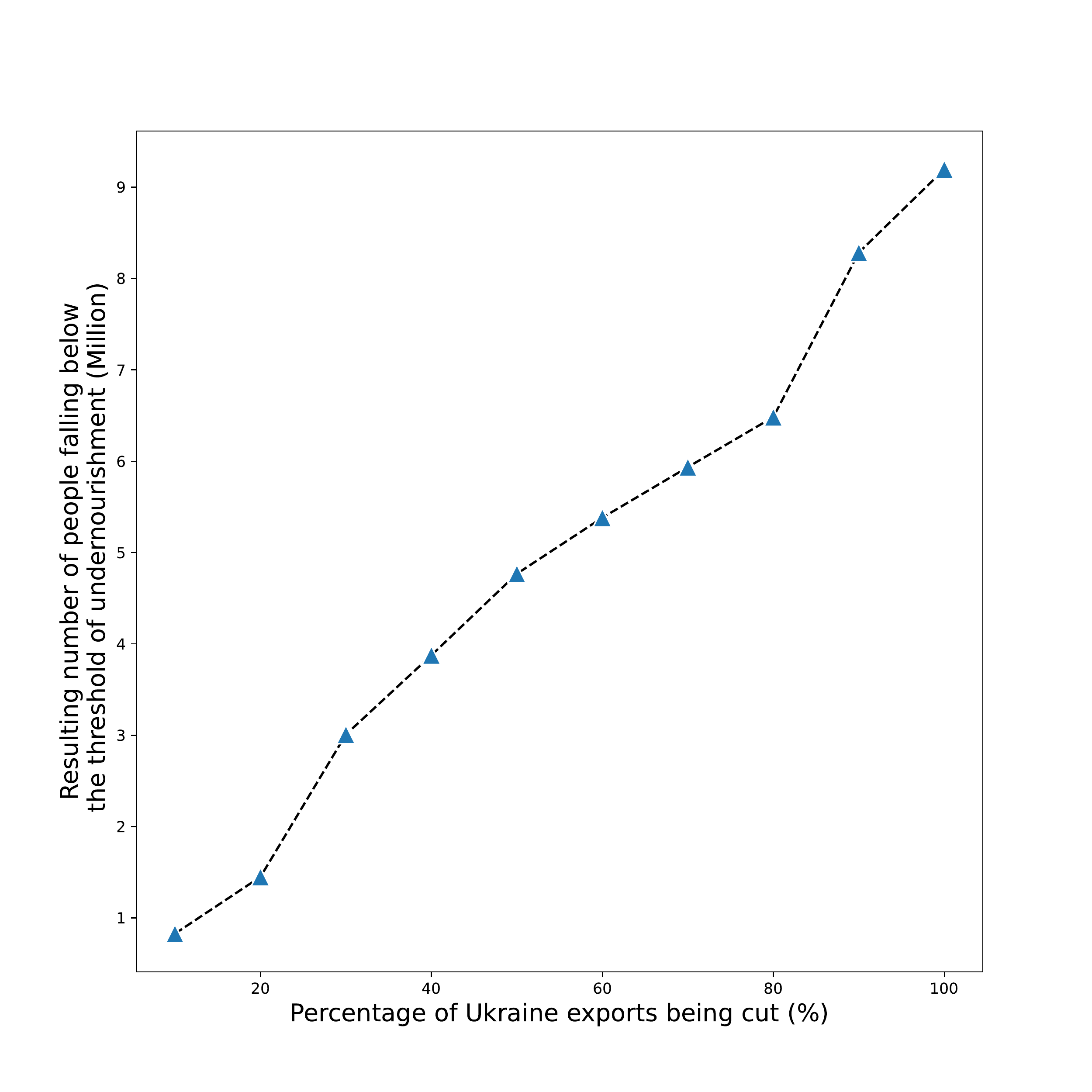}
	\caption{Number of additional undernourished individuals versus the percentage of export drops from Ukraine.}
	\label{fig:variable}
\end{figure}

We redo the simulation for various levels of export drops from $10\%$ up to $100\%$ and the results are shown in Fig.~\ref{fig:variable}.

\section*{Acknowledgments}
JCdG thanks the CLAND for financial support. This work benefited from the support of the French National Research Agency under the `Investissements d’avenir' program with the reference ANR-16-CONV-0003 (CLAND project).

\bibliographystyle{prsty}

\begin{thebibliography}{99}

\bibitem{erenstein2022global}
  Erenstein, O., Jaleta, M., Mottaleb, K.A., Sonder, K., Donovan, J.  Braun, H.-J., Global Trends in Wheat Production,
  Consumption and Trade. Wheat Improvement, 47--66 (2022), Springer.

\bibitem{fao_data}
  Faostat data. FAO. License: CC BY-NC-SA 3.0 IGO.
  Date of access: 07-07-2022. \url{https://www.fao.org/faostat/en/#data}.

\bibitem{Alexander:2006}
Alexander, L.V. et al., Global observed changes in daily climate extremes of temperature and precipitation, Journal of Geophysical Research: Atmospheres, 111 (2006).

\bibitem{Dai:2011}
  Dai, A., Drought under global warming: a review, Wiley Interdisciplinary Reviews: Climate Change, 2:45-65 (2011)

\bibitem{Sheffield:2014}
  Sheffield, J. et al., A drought monitoring and forecasting system for sub-Sahara African water resources and food security,
  Bulletin of the American Meteorological Society, 95:861-882 (2014).

\bibitem{suweis2015resilience}
  Suweis, S., Carr, J., Maritan, A., Rinaldo, A., D’Odorico, P., Resilience and reactivity of global food security, Proceedings of the National Academy of Sciences, 112:6902-6907 (2015).

\bibitem{Portner:2022}
  Portner, H. et al., Climate change 2022: impacts, adaptation and vulnerability, IPCC (2022).

 \bibitem{Hassen}
   Ben Hassen T, El Bilali H. Impacts of the Russia-Ukraine War on Global Food Security:
   Towards More Sustainable and Resilient Food Systems? Foods. 2022; 11(15):2301.

\bibitem{Fagiolo}
  Schweitzer F, Fagiolo G, Sornette D, Vega-Redondo F, Vespignani A,
  White DR. Economic networks: The new challenges. science. 2009 Jul 24;325(5939):422-5.
   
\bibitem{fao:foodBalance}
	Food Balances (2010-). FAO. License: CC BY-NC-SA 3.0 IGO. Date of access: 07-07-2022. \url{https://www.fao.org/faostat/en/#data/FBS}.
	
\bibitem{fao:tradeMatrix}
	Detailed Trade Matrix. FAO. License: CC BY-NC-SA 3.0 IGO. Date of access: 07-07-2022. \url{https://www.fao.org/faostat/en/#data/TM}.

\bibitem{Lin}
  Lin et al. The impact of Russia-Ukraine conflict on global food security, Global Food Security,
  Volume 36, March 2023, 100661


\bibitem{tamea2016global}
  Tamea, S., Laio, F., Ridolfi, L., Global effects of local food-production crises: a virtual water perspective,
  Scientific reports, 6:1-14, (2016).

\bibitem{Andreyeva:2010}
  Andreyeva, T., Long, M., Brownell, K.D., The impact of food prices on consumption: a systematic review of research on the price elasticity of demand for food, American journal of public health, 100:216-222 (2010).

\bibitem{Kumar:2011}
  Kumar, P., Kumar, A., Shinoj, P., Raju, SS, Estimation of demand elasticity for food commodities in India,
  Agricultural Economics Research Review, 24:1-14 (2011).

 \bibitem{Bentley:2022}
 Bentley, A. Broken bread — avert global wheat crisis caused by invasion of Ukraine, Nature 603, 551 (2022)
  
\bibitem{fao:foodSecIndic}
          Suite of Food Security Indicators. FAO. License: CC BY-NC-SA 3.0 IGO. Date of access: 07-07-2022. \url{https://www.fao.org/faostat/en/#data/FS}.

     
\bibitem{serrano2003topology}
        Serrano, M. A., Boguná, M. (2003). Topology of the world trade web. Physical Review E, 68(1), 015101.

\bibitem{serrano2007patterns}
        Serrano, M. N., Boguñá, M., Vespignani, A. (2007). Patterns of dominant flows in the world trade web. Journal of Economic Interaction and Coordination, 2(2), 111-124.

\bibitem{garlaschelli2005structure}
        Garlaschelli, D., Loffredo, M. I. (2005). Structure and evolution of the world trade network. Physica A: Statistical Mechanics and its Applications, 355(1), 138-144.

\bibitem{gutierrez2021analysis}
Gutiérrez-Moya, E., Adenso-Díaz, B., Lozano, S. (2021). Analysis and vulnerability of the international wheat trade network. Food security, 13(1), 113-128.

\bibitem{Deltas:2003}
          Deltas, G. (2003). The small-sample bias of the Gini coefficient: results and implications for empirical research. Review of economics and statistics, 85(1), 226-234.


\bibitem{Watts:1998}
	Watts, D. J., \& Strogatz, S. H. (1998). Collective dynamics of ‘small-world’networks. nature, 393(6684), 440-442.
	
\bibitem{Barabasi:2003}
	Barabási, A. L., \& Bonabeau, E. (2003). Scale-free networks. Scientific american, 288(5), 60-69.

\bibitem{Distefano:2018}
	Distefano, T., Laio, F., Ridolfi, L., \& Schiavo, S. (2018). Shock transmission in the international food trade network. PloS one, 13(8), e0200639.  



\end{thebibliography}

\end{document}